\documentclass[11pt,aps,nofootinbib,showpacs,showkeys,preprintnumbers,amsmath,amssymb,floatfix,11pt]{revtex4-1}
\usepackage[latin9]{inputenc}
\usepackage{graphicx}
\usepackage{esint}



\begin{document}

\title{Resonance model for non-perturbative inputs to gluon distributions
in the hadrons}

\author{B.I.~Ermolaev}

\affiliation{Ioffe Physico-Technical Institute, 194021 St.Petersburg, Russia}

\author{F.~Olness}

\affiliation{Southern Methodist University, Dallas, TX 75275, USA}

\author{S.I.~Troyan}

\affiliation{St.Petersburg Institute of Nuclear Physics, 188300 Gatchina, Russia}

\date{\today}
\begin{abstract}
We construct non-perturbative inputs for the elastic gluon-hadron
scattering amplitudes in the forward kinematic region for both polarized
and non-polarized hadrons. We use the optical theorem to relate invariant
scattering amplitudes to the gluon distributions in the hadrons. By
analyzing the structure of the UV and IR divergences, we can determine
theoretical conditions on the non-perturbative inputs, and use these
to construct the results in a generalized Basic Factorization framework
using a simple Resonance Model. These results can then be related
to the $K_{T}$ and Collinear Factorization expressions, and the corresponding
constrains can be extracted. 
\end{abstract}


\maketitle

\section{Introduction}

The description of hadronic reactions at high energies requires the
use of Quantum Chromodynamics (QCD) in both the perturbative and non-perturbative
domains; such calculations are challenging because the non-perturbative
characteristics of QCD are difficult to quantify. The standard approach
is to use the QCD factorization to divide the problem into perturbative
and non-perturbative components, and then use the properties of the
perturbative expressions to infer basic features of the non-perturbative
piece. We will make use of these properties, together with a simple
Resonance Model, to characterize the non-perturbative inputs that
enter the hadronic scattering process. 

\begin{figure}
\includegraphics[width=0.15\textwidth]{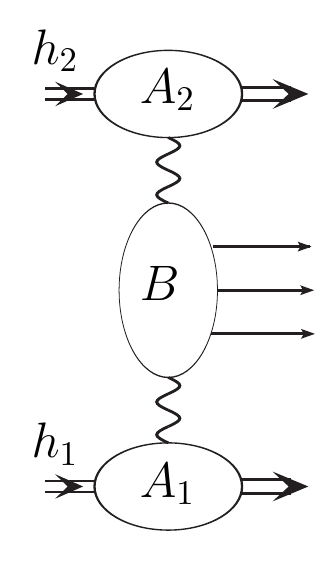} \caption{\label{fig:1} Amplitude for two hadrons ($h_{1}$ and $h_{2}$) with
a single parton (gluon) exchange. Blobs $A_{1,2}$ denote the hadrons
from which the partons are emitted, and blob B depicts the parton
interactions, with the outgoing arrows denoting the produced partons.
The outgoing double arrows on blobs $A_{1,2}$ stand for the final
state spectators.}
\end{figure}

Within the QCD factorization framework, the $2\to n$ hadronic scattering
process (for a single parton exchange%
\footnote{To be specific, we will consider only the case of single-parton collisions;
an overview of double-parton collisions together with an extensive
bibliography can be found in Ref.~\cite{szczur}.%
}) is depicted in Fig.~\ref{fig:1} where we see the gluon exchange
between the hadronic blobs $A_{1,2}$ and the hard scattering process
represented by $B$. The remaining partons of hadrons $h_{1,2,}$
which do not participate in the hard interaction are spectators, and
they are represented by the outgoing double arrows. 

\begin{figure}
\includegraphics[width=0.35\textwidth]{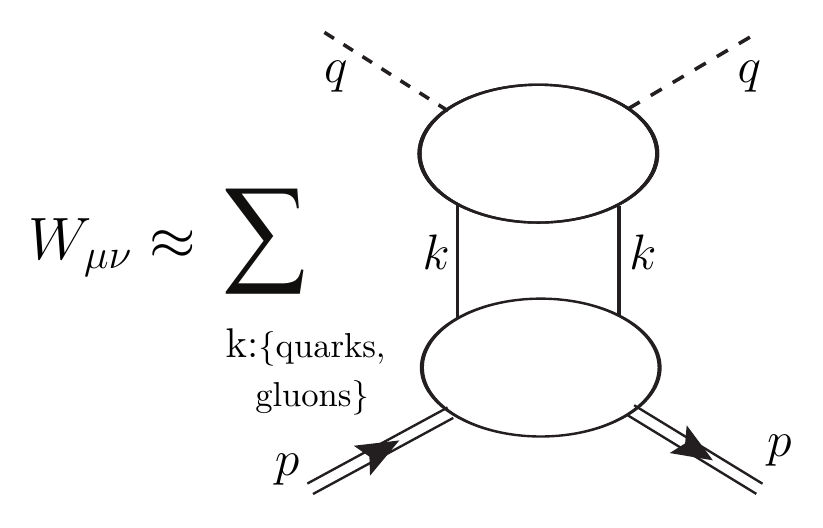} \caption{\label{fig:2} The QCD factorization for the DIS hadronic tensor $W_{\mu\nu}$.
The lowest blob includes the non-perturbative input (PDFs), while
the upper blob corresponds to DIS of the active parton. If we insert
a (vertical) on-shell cut in this diagram (s-cut), it represents a
squared amplitude $AA^{*}\sim d\sigma$.  Without a cut, it represents
an amplitude for a two-parton exchange. }
\end{figure}

The cross section is related to the square of the factorized amplitudes.
We can represent this diagrammatically by combining Fig.~\ref{fig:1}
together with its mirror image (representing the conjugate amplitude).
For the case of Deeply Inelastic Scattering (DIS) where we only have
a single hadron, this is depicted in Fig.~\ref{fig:2} which shows
an incoming hadron ($p$) and photon ($q$) which exchange an intermediate
parton ($k$). Here, the lower blob represents the non-perturbative
input (Parton Distribution Function) describing the emission/absorption
of the exchanged parton $k$ from the initial hadron $p$, and the
upper blob corresponds to the interaction between parton $k$ and
the incident photon $q$. There is an implied (but not drawn) vertical
s-channel cut through this diagram which would represent the DIS on-shell
final state of the process. In total, this diagram then represents
the DIS hadronic tensor $W^{\mu\mbox{\ensuremath{\nu}}}$ which, when
combined with the leptonic tensor $L_{\mu\nu}$, yields the cross
section: $d\sigma\sim L_{\mu\nu}W^{\mu\nu}$.


However, without the vertical s-channel cut Fig.~\ref{fig:2} can
be interpreted as an amplitude $A_{\mu\nu}$ with two partons (each
of momentum $k$) being exchanged in the t-channel; essentially, this
becomes the elastic Compton scattering process. The optical theorem
relates the imaginary part $(\Im)$ of the scattering amplitudes with
the cross section, and we can make use of this to relate amplitude
$A_{\mu\nu}$ to the hadronic tensor (proportional to the cross section)
$W_{\mu\nu}$ as follows:

\begin{equation}
W_{\mu\nu}=\frac{1}{\pi}\ \Im\, A_{\mu\nu}.\label{opt}
\end{equation}
Thus we can compute the cross section for the single-exchange DIS
process ($d\sigma\sim L_{\mu\nu}W^{\mu\nu}$) using the amplitude
for the double-exchange Compton amplitude $A_{\mu\nu}$. In order
to avoid misunderstanding, we note that $t$-channel intermediate
states in the upper (perturbative) blob can involve unlimited number
of partons, even though the blob stands for the $2\to2$ scattering
amplitude. Throughout the paper we will refer to such blobs as non-perturbative
inputs regardless of whether they have an s-channel cut or not.

There are a variety of perturbative QCD factorization frameworks in
the literature, and each is tailored to a specific purpose. For example,
the \textbf{DGLAP} factorization\cite{dglap} and its generalizations\cite{egtg1sum}
describe the case of collinear on-shell partons exchanged in the hadronic
tensor $W_{\mu\nu}$. Correspondingly, one therefore needs on-shell
non-perturbative inputs (i.e., PDFs) for this calculation; such inputs
are the subject of \textbf{Collinear Factorization}\cite{colfact}.
In contrast, the \textbf{BFKL} factorization\cite{bfkl} operates
with essentially off-shell external partons; this precludes a simple
matching with the above collinear factorization. Instead, the \textbf{$K_{T}$-Factorization}\cite{ktfact}
(also referred to as \textbf{High-Energy Factorization}\cite{hefact})
can provide a link to the BFKL framework. 

These factorizations use different parametrizations for the momentum
of the exchanged parton $k$ depending upon which kinematics they
wish to emphasize. For example, in Collinear Factorization, we assume
the parton $k$ is collinear to the hadron $p$, so we use the single
variable $\beta$ to parameterize this relation: 
\begin{equation}
k=\beta p.\label{kcol}
\end{equation}
Thus, $\beta$ is the longitudinal momentum fraction of the parton. 

For $K_{T}$-Factorization, the exchanged parton can be off-shell
and have a transverse momentum $k_{\perp}$ relative to the hadron
momentum $p$, in addition to the longitudinal momentum. We thus use
the parametrization:

\begin{equation}
k=\beta p+k_{\perp},\label{kkt}
\end{equation}
with transverse momentum $k_{\perp}$ parametrizing the two-dimensional
transverse space, and as before $\beta$ parametrizes the longitudinal
space.

While Eq.~(\ref{kkt}) is more general than Eq.~(\ref{kcol}), we
can generalize even further using the standard Sudakov representation\cite{sud}
which involves two longitudinal and two transverse parameters:

\begin{equation}
k=-\alpha p'+\beta q'+k_{\perp}.\label{sud}
\end{equation}
Here, the light-cone momenta $p',q'$ are comprised of the external
momenta $p$ and $q$ as follows: 
\begin{equation}
p'=p-x_{p}q,\quad q'=q-x_{q}p,\quad x_{p}=p^{2}/w,\quad x_{q}=q^{2}/w,\quad w=2p'q'\approx2pq\ .\label{pq}
\end{equation}
They satisfy the inequality $|pq|\gg|p^{2}|,|q^{2}|$ (cf., Fig.~\ref{fig:2}).

In Ref.~\cite{egtfact} we presented the general factorization form
(Basic Factorization) which parametrizes all components of the parton
momentum $k$, and this can systematically be related to both $K_{T}$-Factorization
and to Collinear Factorization. In the literature, $K_{T}$-Factorization
and Collinear Factorization QCD operate with totally different non-perturbative
inputs. The Collinear Factorization makes use of the common PDFs which
are a function of the parton momentum fraction $x$, while the $K_{T}$-Factorization
uses a more generalized non-perturbative object which depends both
on $x$ and the $k_{\perp}$ of the parton. While the differences
stem from the details of the intended application, analyzing them
within the Basic Factorization framework allows us to apply common
theoretical constraints which can be derived from the analysis of
the infra-red (IR) and ultra-violet (UV) singularities of the factorization
convolutions.\cite{egtfact} Because a physical cross section must
be free of any IR and UV cut-offs, we can deduce the properties of
the non-perturbative inputs by imposing the condition that these singularities
must cancel in the total result. 

To further investigate the general case of the Basic Factorization,
we will use the Resonance Model as outlined in Ref.~\cite{egtquark}.
This model is based on the observation that after a hadron emits a
quark or gluon parton, the hadron remnants are unstable and will decay
into a number ($n>1$) of resonant states. Ref.~\cite{egtquark}
examined the case for quarks, and here we extend this to the case
of gluons. We will consider both polarized and unpolarized hadrons,
and relate the above general case to both the $K_{T}$ and Collinear
Factorizations. Our paper is organized as follows: In Section~II
we evaluate the elastic gluon-hadron amplitudes for the forward kinematic
region in the Born approximation, and then analyze the impact of radiative
corrections. We investigate the convergence of the the factorization
convolutions to determine the the related constraints on the non-perturbative
inputs. We then use those restrictions in Section~III within our
Resonance Model to construct the non-perturbative inputs for the Basic
Factorization, and then derive the corresponding results in the $K_{T}$-
and Collinear Factorizations; we consider both polarized and unpolarized
gluons. We also compare the results for both the $K_{T}$- and Collinear
Factorizations with the commonly used formula from the literature.
In Section~IV we discuss these results and provide an outlook.

\section{Elastic gluon-hadron scattering amplitudes with forward kinematics}

\begin{figure}
\includegraphics[width=0.25\textwidth]{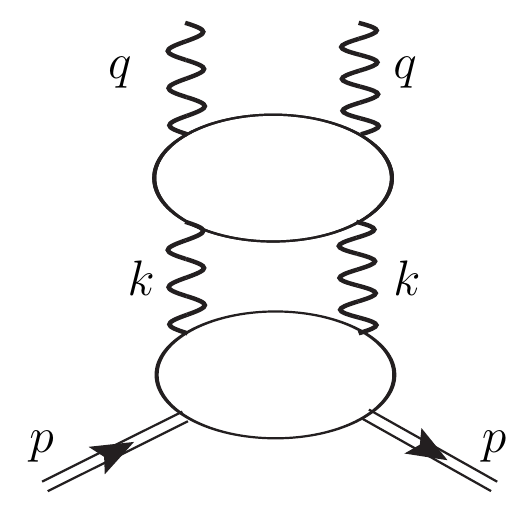} \caption{\label{fig:3} The amplitude factorization of hadron-gluon scattering
in the forward kinematic region, with the exchange of intermediate
partons (gluons). The upper blob ($H$) depicts $2\to2$ gluon scattering,
and the lower blob ($T$) contains the hadronic target. }
\end{figure}

In this section we study the elastic gluon-hadron amplitudes in the
forward kinematic region for the case of an intermediate gluon. We
start with the Basic Factorization framework, and find the conditions
for the factorization convolutions to be free of both UV and IR singularities;
these restriction will help us model the non-perturbative inputs. 

The elastic gluon-hadron scattering amplitude amplitude $A$ receives
contributions from both an s-channel and u-channel process;%
\footnote{We consider the $t$-channel color singlets only.%
} the s-channel is depicted in Fig.~\ref{fig:3} and u-channel can
be obtained by the replacement $q\to-q$. For this particular set
of graphs, the imaginary part ($\Im$) vanishes, so it does not contribute
to the gluon distribution.

\subsection{Gluon-hadron scattering amplitudes in the Born approximation}

\begin{figure}
\includegraphics[width=0.25\textwidth]{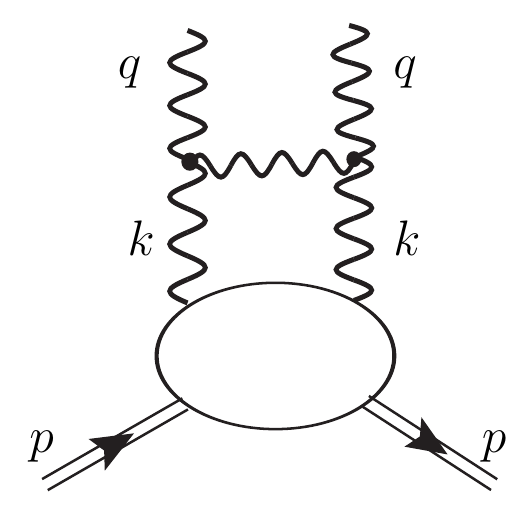} \caption{\label{fig:4} The amplitude factorization of hadron-gluon scattering
in the Born approximation, \emph{c.f.}, Fig.~\ref{fig:3}}
\end{figure}

In the Basic Factorization framework, the upper blob of Fig.~\ref{fig:3}
is perturbative while the lower blob includes only non-perturbative
contributions. If we work in the Born approximation, the leading contribution
for Fig.~\ref{fig:3} is a single gluon exchange as shown in Fig.~\ref{fig:4}.
We can use the standard Feynman rules to obtain an analytic expression
for the elastic gluon-hadron amplitude $A^{B}$ with non-zero imaginary
part $\Im$ in the Born approximation:%
\footnote{Throughout the paper we use the Feynman gauge.%
}

\begin{equation}
A^{B}=\frac{4\pi\alpha_{s}N}{(2\pi)^{4}}\int d^{4}k\ l_{\mu}l'_{\nu}\,\left[\frac{H_{\mu\nu\lambda\rho}(q,k)}{s_{2}+\imath\epsilon}+\frac{H_{\nu\mu\lambda\rho}(-q,k)}{u_{2}+\imath\epsilon}\right]\frac{1}{k^{2}k^{2}}\, T_{\lambda\rho}(k,p,S),\label{bg}
\end{equation}
where $s_{2}=(q+k)^{2}$, $u_{2}=(q-k)^{2}$, $N=3$ is the color
factor and $S$ denotes the hadron spin. The term in squared brackets
corresponds to the Born amplitude for gluon-gluon scattering, $l_{\mu}(q)$
and $l'_{\nu}(q)$ denote the polarization vectors of the external
gluons, and the $1/k^{2}$ terms come from the gluon propagators.
The contributions from the s-channel and u-channel processes are evident
from the $1/(s_{2}+i\epsilon)$ and $1/(u_{2}+i\epsilon)$ terms.
The target function $T_{\lambda\rho}$ contains only non-perturbative
contributions; it corresponds to the lower blob of Fig.~\ref{fig:3}
Both $T_{\lambda\rho}$ and the term in the squared brackets are dimensionless.
The hard perturbative term $H_{\mu\nu\lambda\rho}$ is given by:

\begin{equation}
H_{\mu\nu\lambda\rho}=\left[(-q-2k)_{\mu}g_{\lambda\sigma}+(2q+k)_{\lambda}g_{\mu\sigma}+(k-q)_{\sigma}g_{\mu\lambda}\right]\left[(q+2k)_{\nu}g_{\rho\sigma}+(-2q-k)_{\rho}g_{\nu\sigma}+(q-k)_{\sigma}g_{\nu\rho}\right]\quad.\label{h}
\end{equation}
$H_{\mu\nu\lambda\rho}$ corresponds to the upper blob of Fig.~\ref{fig:3}.

For the target function $T_{\lambda\rho}$ we can write the general
tensor structure as: 
\begin{equation}
T_{\lambda\rho}^{(gen)}=g_{\lambda\rho}A+p_{\lambda}p_{\rho}B+(p_{\lambda}k_{\rho}+k_{\lambda}p_{\rho})C+k_{\lambda}k_{\rho}D\quad.,\label{tgen}
\end{equation}
where $T_{\lambda\rho}^{(gen)}$ is a function of the relevant momenta
$\{p,k\}$ and four arbitrary invariant amplitudes $\{A,B,C,D\}$.
If we were to replace the incoming hadron by a bare quark, then the
non-perturbative target function $T_{\lambda\rho}$ is replaced by
the perturbative quark amplitude $Q_{\lambda\rho}$. This can be decomposed
into the unpolarized part $Q_{\lambda\rho}^{U}$ and the spin-dependent
part $Q_{\lambda\rho}^{S}$:

\begin{equation}
Q_{\lambda\rho}=Q_{\lambda\rho}^{U}+Q_{\lambda\rho}^{S},\label{qussum}
\end{equation}
with 
\begin{eqnarray}
Q_{\lambda\rho}^{U} & = & \left(2p_{\mu}p_{\nu}-k_{\mu}p_{\nu}+p_{\mu}k_{\nu}-pk~g_{\mu\nu}\right)\left(\frac{-8\pi\alpha_{s}C_{F}}{(p-k)^{2}-m_{q}^{2}+\imath\epsilon}\right),\label{qus}\\
Q_{\lambda\rho}^{S} & = & \imath m_{q}\epsilon_{\mu\nu\lambda\rho}k_{\lambda}(S_{q})_{\rho}\left(\frac{-8\pi\alpha_{s}C_{F}}{(p-k)^{2}-m_{q}^{2}+\imath\epsilon}\right),\nonumber 
\end{eqnarray}
where $m_{q}$ is the quark mass, $S_{q}$ is the quark spin and $C_{F}=4/3$. 

We will make the assumption that $T_{\lambda\rho}$ keeps the polarization
structure of $Q_{\lambda\rho}$ so that: 

\begin{equation}
T_{\lambda\rho}=T_{\lambda\rho}^{U}+T_{\lambda\rho}^{S},\label{tussum}
\end{equation}
with

\begin{eqnarray}
T_{\lambda\rho}^{U} & = & \left(2p_{\mu}p_{\nu}-k_{\mu}p_{\nu}-p_{\mu}k_{\nu}+pk~g_{\mu\nu}\right)\ M_{U}(s_{1},k^{2}),\label{tus}\\
T_{\lambda\rho}^{S} & = & \imath\, M_{h}\epsilon_{\mu\nu\lambda\rho}k_{\lambda}S_{\rho}\ M_{S}(s_{1},k^{2}),\nonumber 
\end{eqnarray}
where $s_{1}=(p-k)^{2}$, and $M_{h}$ and $S$ are the hadron mass
and spin, respectively.%
\footnote{While the unpolarized and spin-dependent quark amplitudes $Q^{U}$
and $Q^{S}$ had a common invariant factor (c.f., Eq.~(\ref{qus})),
there is no assumption that the invariant amplitudes $M_{U}$ and
$M_{S}$ coincide. %
} Substituting $T_{\lambda\rho}^{U,S}$ into the elastic gluon-hadron
amplitude of Eq.~(\ref{bg}), we obtain:

\begin{equation}
A_{U}=\frac{4\pi\alpha_{s}N}{(2\pi)^{4}}\int\frac{d^{4}k}{k^{2}k^{2}}\left[\frac{N_{s}^{(1)}}{s_{2}+\imath\epsilon}+\frac{N_{u}^{(1)}}{u_{2}+\imath\epsilon}\right]M_{U}(s_{1},k^{2}),\label{au}
\end{equation}

\begin{equation}
A_{S}=\frac{4\pi\alpha_{s}N}{(2\pi)^{4}}\int\frac{d^{4}k}{k^{2}k^{2}}\left[\frac{N_{s}^{(2)}}{s_{2}+\imath\epsilon}+\frac{N_{u}^{(2)}}{u_{2}+\imath\epsilon}\right]M_{S}(s_{1},k^{2}),\label{as}
\end{equation}
with

\begin{eqnarray}
N_{s}^{(1)} & = & 4k^{2}(2m_{h}^{2}+2pk)-16(2(pq)^{2}+2(pq)(pk)-q^{2}(pk))+u_{2}(2M_{h}^{2}-2pk)\ ,\label{n12}\\
N_{s}^{(2)} & = & \imath M_{h}\epsilon_{\lambda\rho\sigma\tau}S_{\tau}\left[8q_{\rho}\left(l'_{\lambda}k_{\sigma}(kl)-l_{\lambda}k_{\sigma}(kl')\right)-u_{2}k_{\sigma}l_{\lambda}l'_{\rho}\right]\ .\nonumber 
\end{eqnarray}
For the unpolarized amplitude $A^{U}$, we have summed over the gluon
polarizations in the expression for $N_{s}^{(1)}$.

\subsection{Analysis of IR and UV singularities }

We now examine the IR and UV singularity structure of the $A_{U,S}$
amplitudes. As these quantities are related to physical cross sections,
they must ultimately be finite; therefore, the singularities must
cancel. In what follows, we will find it convenient to use the Sudakov
variables of Eq.~(\ref{sud}); specifically, we have: 

\begin{equation}
2pk=-\alpha w+\beta x_{p}w,\quad2qk=\beta w-\alpha xw,\quad k^{2}=-\alpha\beta w+k_{\perp}^{2}\ .\label{sudinv}
\end{equation}

The gluon propagators give rise to the factors $1/k^{2}$, and this
will lead to an IR singularity $k^{2}=0$. If we were to introduce
an IR cut-off, the result then depends on this unphysical parameter
and it must be canceled in the final physical cross section. For the
amplitude $A_{U,S}$, which we will relate to a physical cross section
using the optical theorem, we are thus unable to introduce any IR
cut-off. Therefore, our only option is that the $M_{U,S}$ amplitudes
in Eqs.~(\ref{au},\ref{as}) must cancel the IR singularities. In
order to compensate the $1/(k^{2}k^{2})$ gluon propagators of Eqs.~(\ref{au},\ref{as}),
in the limit $k^{2}\to0$ $M_{U,S}$ should obey:

\begin{equation}
M_{U,S}(s_{1},k^{2})\sim\left(k^{2}\right)^{1+\eta}\label{mir}
\end{equation}
with $\eta>0$.

We also encounter UV singularities in Eqs.~(\ref{au},\ref{as})
at large $|k|$, or equivalently (in terms of the Sudakov variables)
at large $|\alpha|$ where the integrands of Eqs.~(\ref{au},\ref{as})
behave as:

\begin{equation}
\sim\frac{\alpha^{2}}{\alpha^{3}}\ M_{U,S}\ \equiv\frac{1}{\alpha}\ M_{U,S}\ .\label{uv}
\end{equation}
In order to guarantee the amplitudes $A_{U,S}$ are UV finite then
$M_{U,S}$ should decrease at large $\alpha$ as:

\begin{equation}
M_{U,S}\sim\alpha^{-\chi},\label{muv}
\end{equation}
with $\chi>0$. We note that the restrictions in Eqs.~(\ref{mir},\ref{muv})
apply to the most general polarization structure of $T_{\lambda\rho}$
as given in Eq.~(\ref{tgen}).

\subsection{Gluon-hadron scattering amplitudes beyond the Born approximation}

It is relatively easy to extend the above Born result to include radiative
corrections. Pictorially, the single gluon exchange of Fig.~\ref{fig:4}
becomes the generic upper blob of Fig.~\ref{fig:3} which now includes
the higher-order loop contributions which may be divergent. We now
consider the general types of divergences which may enter, and assess
whether the radiative corrections will modify the Born restrictions
of Eqs.~(\ref{mir},\ref{muv}).

We first consider the case where the upper blob of Fig.~\ref{fig:3}
acquires additional divergences due to the radiative corrections.
As QCD is renormalizable, the UV divergences are absorbed into the
redefinition/renormalization of the QCD couplings and masses; there
are no complications here. For the IR divergences, these can be regulated
with a cut-off such as the parton virtuality $k^{2}$. These procedures
will render the amplitude finite.

The upper blob of Fig.~\ref{fig:3} represents the generic $2\to2$
amplitude and can depend only on the total energy $(q+k)^{2}=q^{2}+\omega\beta+k^{2}$
and virtualities $q^{2}$ and $k^{2}$. The only IR divergent terms
which can appear at higher order are logarithms such as $\ln(s\beta/k^{2})$;
here, $k^{2}$ acts as an IR cut-off, and this logarithmic divergence
will not alter the constraint of Eq.~(\ref{mir}). Concerning the
UV divergences, the upper blob depends on $\alpha$ only through $k^{2}$,
so any new $\alpha$-dependent divergences from radiative corrections
are also of the logarithmic type; in a similar manner, these divergences
will not alter the constraint of Eq.~(\ref{muv}). 

Let us note that the situation with UV divergences in the conventional
factorization approach can be more complex; for example the analysis
of UV divergences for $K_{T}$ -factorization is complicated by the
existence of rapidity gaps.%
\footnote{This problem was first considered in Ref.~\cite{collinsrapid} and
then in Ref.~\cite{cheredrapid}. An overview of the rapidity gaps
can be found in the Ref.~\cite{sterm}. %
} The essence of the problem is that the perturbative contributions
are divided between the upper and lower blob of Fig.~\ref{fig:3},
and this complicates the cancellation of the UV divergences. The IR
divergences are more delicate and must be regulated individually with
either a cut-off or a gluon mass regulator; the latter can be easily
achieved by keeping the external gluons off-shell so their virtualities
act as IR cut-offs for integrations of the loops.

\section{Modeling the non-perturbative invariant amplitudes $M_{U,S}$}

We will now characterize the structure of the non-perturbative inputs
to the parton distributions (both the usual and generalized) in the
context of the Resonance Model. This was outlined in Ref.~\cite{egtquark}
for quark distributions, and we extend it to describe the gluon distributions
in the hadrons. We begin by applying this model to the non-perturbative
invariant amplitudes $M_{U,S}$ in the Basic Factorization framework,
and later we will relate this to the usual $K_{T}$- and Collinear
Factorization results. 

We review the criteria that invariant amplitudes $M_{U,S}$ should
satisfy: 
\begin{description}
\item [{Criterion}] \textbf{(i)} $M_{U,S}$ should satisfy the requirements
of Eqs.~(\ref{mir},\ref{muv}) so that the physical results are
IR and UV finite.
\item [{Criterion}] \textbf{(ii)} $M_{U,S}$ should have a non-zero imaginary
parts in $s_{1}$ to facilitate use of the optical theorem which relates
the elastic gluon-hadron amplitudes to gluon distributions in the
hadrons. 
\item [{Criterion}] \textbf{(iii)} $M_{U,S}$ should allow for the step-by-step
reduction of the Basic Factorization results to those of the $K_{T}$-
and Collinear Factorization results.%
\footnote{Such a reduction was suggested in Ref.~\cite{egtfact}.%
} 
\end{description}

\subsection{Non-perturbative gluon inputs for Basic Factorization}

As a first step, we posit that we can decompose the amplitudes $M_{U,S}(s_{1},k^{2})$
into independent functions of the separate momenta $\{s_{1},k^{2}\}$:

\begin{equation}
M_{U}(s_{1},k^{2})=R_{U}\left(k^{2}\right)\ Z_{U}\left(s_{1}\right),\qquad M_{S}(s_{1},k^{2})=R_{S}\left(k^{2}\right)\ Z_{S}\left(s_{1}\right)\ .\label{musgen}
\end{equation}
In the following we will manipulate $M_{U}$ and $M_{S}$ in parallel,
so we drop the subscripts in the following and use: $M(s_{1},k^{2})=R(k^{2})\ Z(s_{1})$.
To satisfy the IR constraints of Eq.~(\ref{mir}), we find at small
$k^{2}$ that $R$ should behave as:

\begin{equation}
R\sim\left(k^{2}\right)^{1+\eta}\label{rsmallk}
\end{equation}

Similarly, the UV constraints of Eq.~(\ref{muv}) impose the condition:
\[
M(s_{1},k^{2})=R(k^{2})\ Z(s_{1})\sim\alpha^{-\chi}
\]
 at large $|\alpha|$. While the behavior of $R$ at large $|\alpha|$
is ambiguous, it could be that the small-$k^{2}$ behavior at large
$|\alpha|$ is again $R\sim(k^{2})^{1+\eta}$. Expressing this in
Sudakov variables according to Eq.~(\ref{sudinv}) we have:

\begin{equation}
R\sim\alpha^{1+\eta}.\label{rbiga}
\end{equation}
 Since this is the most UV divergent case, we can use this together
with Eqs.~(\ref{muv},\ref{rbiga}) to conclude that $Z$ should
behave as: 

\begin{equation}
Z\sim\alpha^{-1-\eta-\chi}\ ,\label{zuv}
\end{equation}
at large $|\alpha|$

We will now construct a $Z$ satisfying Eq.~(\ref{zuv}) and make
use of our Resonance Model. This is based on the idea that after the
initial hadron has emitted a parton,%
\footnote{It does not matter whether the emitted parton is a quark or a gluon;
the important observation is that it is a colored object. %
} the hadron remnant has unbalanced colors and cannot be stable. As
we hypothesize that this unstable state will decay into a number of
resonances, we then take $Z$ to be a product of Breit-Wigner functions: 

\begin{equation}
Z(s_{1})\approx Z_{n}(s_{1})=\prod_{r=1}^{n}\frac{1}{\left(s_{1}-M_{r}^{2}+\imath\Gamma_{r}\right)},\label{zgen}
\end{equation}
where $n>1$ (since it must decay into multiple resonances). For the
sake of simplicity, we consider the minimum allowable value $n=2$
and approximate $Z(s_{1})$ as an interference of two resonances:

\begin{equation}
Z(s_{1})=\frac{1}{\left(\Delta M_{12}^{2}+\imath\Delta\Gamma_{12}\right)}\left[\frac{1}{\left(s_{1}-M_{1}^{2}+\imath\Gamma_{1}\right)}-\frac{1}{\left(s_{1}-M_{2}^{2}+\imath\Gamma_{2}\right)}\right],\label{zres}
\end{equation}
with $\Delta M_{12}^{2}=M_{1}^{2}-M_{2}^{2},~\Delta\Gamma_{12}=\Gamma_{1}-\Gamma_{2}$.
In terms of the Sudakov variables, 
\begin{equation}
M(s_{1},k^{2})\equiv M(\alpha,\beta,k^{2})=R(k^{2})\ Z(s_{1})=\frac{R(k^{2})}{C_{Z}}\left[\frac{1}{\left(w\alpha-\mu_{1}^{2}+k^{2}+\imath\Gamma_{1}\right)}-\frac{1}{\left(w\alpha-\mu_{2}^{2}+k^{2}+\imath\Gamma_{2}\right)}\right],\label{msud}
\end{equation}
where $C_{Z}=\left(\Delta M_{12}^{2}+\imath\Delta\Gamma_{12}\right)$
and $\mu_{1,2}^{2}=M_{1,2}^{2}-p^{2}$. Applying the optical theorem
to Eq.~(\ref{msud}) allows us to obtain the non-perturbative contribution
$\Psi$ to the gluon distributions in the hadrons: 

\begin{equation}
\Psi=-\Im\, T=\widetilde{R}(k^{2})\left[\frac{\Gamma_{1}}{\left(w\alpha+k^{2}-\mu_{1}^{2}\right)^{2}+\Gamma_{1}^{2}}-\frac{\Gamma_{2}}{\left(w\alpha+k^{2}-\mu_{2}^{2}\right)^{2}+\Gamma_{2}^{2}}\right].\label{psi}
\end{equation}
Obviously, the expression $\Psi$ is of the Breit-Wigner type because
this was the form used in the \textit{ansatz} for our Resonance Model
of Eq.~(\ref{zgen}).

\subsection{Non-perturbative gluon inputs in $K_{T}$- Factorization}

The expression in Eq.~(\ref{msud}) for the non-perturbative input
$M(\alpha,\beta,k^{2})$ is obtained in the Basic Factorization framework.
As described in Ref.~\cite{egtfact,egtquark}, we can relate this
to $K_{T}$-Factorization by integrating out the $\alpha$ variable
to obtain $M_{KT}\left(\beta,k_{\perp}^{2}\right)$. However, there
is a complication because in Fig.~\ref{fig:3} both the upper and
the lowest blobs depend on $\alpha$, so one cannot integrate $M(\alpha,\beta,k^{2})$
(the lowest blob) over $\alpha$ independently of the upper blob.
Therefore, $M_{KT}(\beta,k_{\perp}^{2})$ cannot be derived from $M(\alpha,\beta,k^{2})$
in a straightforward way.

Nevertheless, we can relate the Basic Factorization and the $K_{T}$-Factorization
in an approximate manner. We first observe that the upper blob depends
on $\alpha$ only through $k^{2}$; if we could limit our integration
to the region where the $\alpha$ dependence of the upper blob is
negligible, then we can effectively integrate out $\alpha$ to obtain
the $K_{T}$-Factorization result; this is our plan. In the integration
region when: 

\begin{equation}
w|\alpha\beta|\ll k_{\perp}^{2}\label{abk}
\end{equation}
the perturbative blobs (the upper blob in Fig.~\ref{fig:3}) is insensitive
to $\alpha$ and the non-perturbative blobs are independent of $\beta$.
Because the upper blob depends on $\alpha$ only through $k^{2}$
in the limit of Eq.~(\ref{abk}), we can use the approximation $k^{2}=-\omega\alpha\beta-k_{\perp}^{2}\approx-k_{\perp}^{2}$.
This makes it possible to integrate $M\left(\alpha,\beta,k^{2}\right)$
independently, and we obtain:

\begin{equation}
M_{KT}(\beta,k_{\perp}^{2})=\int_{-\alpha_{0}}^{\alpha_{0}}d\alpha M(\alpha,\beta,k^{2}),\label{tktgen}
\end{equation}
with

\begin{equation}
\alpha_{0}\ll k_{\perp}^{2}/(w\beta).\label{azero1}
\end{equation}

In Ref.~\cite{egtfact} we discussed a general structure of non-perturbative
inputs in both the Basic and $K_{T}$ -Factorization frameworks, and
we estimated $\alpha_{0}$ to be: 

\begin{equation}
\alpha_{0}=k_{\perp}^{2}/(w\beta).\label{azerobeta}
\end{equation}
Although this estimate adequately describes the main features of the
non-perturbative inputs, it is misleading for detailed quantitative
analysis. Therefore we need to find an improved estimate for $\alpha_{0}$
that agrees with Eq.~(\ref{abk}) and is also independent the variables
associated with the perturbative blob such as $\beta$ and $x$. Given
that $x<\beta<1$, the requirement in Eq.~(\ref{abk}) is satisfied
when

\begin{equation}
\alpha_{0}=\xi\: k_{\perp}^{2},\label{azero}
\end{equation}
with a positive $\xi$ obeying the inequality $\xi\gg1/x$. Combining
Eqs.~(\ref{msud},\ref{tktgen},\ref{azero}) we arrive at the following
expression for the non-perturbative input $M_{KT}(\beta,k_{\perp}^{2})$
in $K_{T}$ -Factorization:

\begin{eqnarray}
M_{KT}(\beta,k_{\perp}^{2}) & \approx & R\left(k_{\perp}^{2}\right)\left[\frac{1}{\xi k_{\perp}^{2}-\mu_{1}^{2}+\imath\Gamma_{1}}+\frac{1}{\xi k_{\perp}^{2}-\mu_{2}^{2}+\imath\Gamma_{2}}+\frac{1}{\xi k_{\perp}^{2}+\mu_{1}^{2}-\imath\Gamma_{1}}+\frac{1}{\xi k_{\perp}^{2}+\mu_{2}^{2}-\imath\Gamma_{2}}\right]\label{tkt}\\
 & = & R'\left(k_{\perp}^{2}\right)\left[\frac{1}{k_{\perp}^{2}-{\mu'}_{1}^{2}+\imath\Gamma'_{1}}+\frac{1}{k_{\perp}^{2}-{\mu'}_{2}^{2}+\imath\Gamma'_{2}}+\frac{1}{k_{\perp}^{2}+{\mu'}_{1}^{2}-\imath\Gamma'_{1}}+\frac{1}{k_{\perp}^{2}+{\mu'}_{2}^{2}-\imath\Gamma'_{2}}\right],\nonumber 
\end{eqnarray}
where $R'=R/\xi,~$, and ${\mu'}_{1,2}^{2}=\mu_{1,2}^{2}/\xi,~\Gamma'_{1,2}=\Gamma_{1,2}/\xi$.
We also used $\xi\pm1\approx\xi$ at large $\xi$. Eq.~(\ref{tkt})
is valid when

\begin{equation}
w\alpha_{0}\gg M_{1,2}^{2},\qquad\Gamma_{1,2}\gg\Delta M^{2},\Delta\Gamma,\label{azeromg}
\end{equation}
that is when $k_{\perp}^{2}$ is away from the resonance region $k_{\perp}^{2}\sim\mu_{1,2}^{2}$.
As $k_{\perp}^{2}$ approaches $\mu_{1,2}^{2}$, the corrections to
Eq.~(\ref{tkt}) will increase. This means that the relation between
the Basic Factorization and the $K_{T}$-Factorization amplitudes
of Ref.~\cite{egtfact} is valid with the Resonance Model outside
the resonance region; inside the resonance region the corrections
can be large, but this can be improved with a redefinition of the
parameters $\mu'_{1,2}$ and $\Gamma'_{1,2}$.

An advantage of Eq.~(\ref{tkt}) is that the resonance form is similar
to Eq.~(\ref{msud}). In order maintain the validity in the resonance
region we choose the following prescription. First, we derive Eq.~(\ref{tkt})
from Eq.~(\ref{msud}) in the limit of Eq.~(\ref{azeromg}) and
then analytically continue it into the resonance region. Such a strategy
is equivalent to independently specifying $M_{KT}$. 

Applying the Optical theorem to Eq.~(\ref{tkt}), we obtain the gluon
distributions $\Phi$ in the $K_{T}$-Factorization framework, where

\begin{equation}
\Phi=\Phi_{R}+\Phi_{B},\label{phirb}
\end{equation}
with

\begin{equation}
\Phi_{R}=R'\left(k_{\perp}^{2}\right)\left(\frac{\Gamma'_{1}}{(k_{\perp}^{2}-{\mu'}_{1}^{2})^{2}+{\Gamma'}_{1}^{2}}+\frac{\Gamma'_{2}}{(k_{\perp}^{2}-{\mu'}_{2}^{2})^{2}+{\Gamma'}_{2}^{2}}\right)\label{phir}
\end{equation}
and

\begin{equation}
\Phi_{B}=-R'\left(k_{\perp}^{2}\right)\left(\frac{\Gamma'_{1}}{(k_{\perp}^{2}+{\mu'}_{1}^{2})^{2}+{\Gamma'}_{1}^{2}}+\frac{\Gamma'_{2}}{(k_{\perp}^{2}+{\mu'}_{2}^{2})^{2}+{\Gamma'}_{2}^{2}}\right).\label{phib}
\end{equation}
We have divided $\Phi$ into resonance ($\Phi_{R}$) and background
($\Phi_{B}$) contributions which are Breit-Wigner forms. The signs
of ${\mu'}_{1}^{2}$ and ${\mu'}_{2}^{2}$ cannot be fixed \textit{a
priori}, but $\Phi_{R}\to\Phi_{B}$ when ${\mu'}_{1,2}^{2}\to-{\mu'}_{1,2}^{2}$,
so we can take ${\mu'}_{1,2}^{2}$ positive without loss of generality.
We recall that $k_{\perp}^{2}>0$, so $\Phi_{R}$ is within the resonant
region $k_{\perp}^{2}\sim\mu_{1,2}^{2}$, while $\Phi_{B}$ is outside
that region. Therefore, the non-perturbative input $\Phi$ in the
$K_{T}$ -Factorization is represented by its resonance part $\Phi_{R}$
and the background contribution $\Phi_{B}$. Despite the overall minus
sign in Eq.~(\ref{phib}), it turns out that the background contribution
is positive because the main contribution to the integral of $\Phi_{B}$
over $k_{\perp}^{2}$ comes from the lower limit.

\subsection{Non-perturbative gluon inputs in Collinear Factorization}

The relation from the non-perturbative input $\Phi(k_{\perp}^{2})$
in $K_{T}$-Factorization to the input $\varphi$ in Collinear Factorization
can be obtained by integrating $\Phi$ over $k_{\perp}$. This approximate
relation is:

\begin{equation}
\varphi({\mu'}_{1}^{2},{\mu'}_{2}^{2})\approx\left[R'\left({\mu'}_{1}^{2}\right)+R'\left({\mu'}_{2}^{2}\right)\right]-\int_{0}^{w}dk_{\perp}^{2}\Phi_{B}\left(k_{\perp}^{2}\right).\label{phi}
\end{equation}
Here, the expression in the squared brackets corresponds to the integration
of $\Phi_{R}$, while integration of $\Phi_{B}$ is the explicit integral. 

We recall that the non-perturbative inputs $\varphi({\mu'}_{1}^{2},{\mu'}_{2}^{2})$
differ significantly from the integrated distributions $\phi(\beta,\mu_{col}^{2})$
conventionally used in the Collinear Factorization. These differences
include the following important points: 
\begin{description}
\item [{(i)}] The non-perturbative inputs $\varphi({\mu'}_{1}^{2},{\mu'}_{2}^{2})$
do not depend on $\beta$, while the conventional integrated parton
distribution $\phi(\beta,\mu_{col}^{2})$ explicitly does depend on
this variable. 
\item [{(ii)}] $\varphi({\mu'}_{1}^{2},{\mu'}_{2}^{2})$ are altogether
non-perturbative while $\phi(\beta,\mu_{col}^{2})$ includes both
perturbative and non-perturbative contributions.%
\footnote{In Ref.~\cite{egtfact} we demonstrated that evolution of $\varphi$
from the scales ${\mu'}_{1,2}^{2}$ to an arbitrary scale $\mu_{col}^{2}$
can be done perturbatively by moving contributions from the perturbative
blobs to $\varphi$.%
} 
\item [{(iii)}] The factorization scale $\mu_{col}$ is arbitrary and usually
is chosen in the perturbative domain ($\mu\sim1$~GeV) while the
non-perturbative scales ${\mu'}_{1,2}^{2}$ are associated with the
maximum of $\Phi(k_{\perp}^{2})$; hence, they cannot be chosen arbitrary
and must either be in the non-perturbative or perturbative domain.
\end{description}

\subsection{Comparisons with the $K_{T}$ and collinear factorization frameworks}

As a final step, we will make a qualitative comparison of $\varphi$
with the standard DGLAP parton distribution. In Eq.~(\ref{phi}),
$\phi$ does not depend on the longitudinal variable $\beta$. This
is in contrast to the parton distributions in the Collinear Factorization
framework which explicitly depend on $\beta$, and this can be parametrized
as:\cite{fits}

\begin{equation}
\phi(\beta,\mu^{2})=N\,\beta^{-a}\ (1-\beta)^{b}\,(1+c\beta^{d})\quad,\label{dglapfitabs}
\end{equation}
where $N$ is a normalization, and the phenomenological parameters
$a,b,c,d>0$ are fit to experimental data at a specific factorization
scale $\mu$. As was demonstrated in Ref.~\cite{egtg1sum}, the $\beta^{-a}$
term of Eq.~(\ref{dglapfitabs}) resums the leading logarithmic contributions;
it should be removed when the resummation is included explicitly.
Similarly, it is suggested (though not proven) that $(1-\beta)^{b}$
resums the $ln^{n}(1-x)$ terms; again, it should be removed when
the resummation is included explicitly. The $(1+c\beta^{d})$ terms
resum the residual contributions, and can be removed when the above
two resummations are performed. The remaining normalization $N$ then
corresponds to the non-perturbative input $\varphi$ obtained in Eq.~(\ref{phi}).

While we were able to use the Resonance Model to suggest a form for
the $Z$ factor, we do not have an analogous model for R; hence, it
is arbitrary up to the restrictions of Eq.~(\ref{rsmallk}). In the
Basic Factorization framework we have $R\sim\left(k^{2}\right)^{1+\eta}$
at small $k^{2}$, so a possible functional form could be:

\begin{equation}
R=\left(\frac{k^{2}}{k^{2}+M^{2}}\right)^{1+\eta}.\label{r}
\end{equation}

In the $K_{T}$-Factorization framework, $k^{2}$ of Eq.~(\ref{r})
is replaced by $k_{\perp}^{2}$. One possible form for R suggested
by Ref.~\cite{zotov} is a Gaussian in $k_{\perp}^{2}$ such as: 

\begin{equation}
R=\left[a_{1}\left(k_{\perp}^{2}\right)^{b_{1}}+a_{2}\left(k_{\perp}^{2}\right)^{b_{2}}\right]\exp\left[-\lambda k_{\perp}^{2}\right]\,,\label{rzot}
\end{equation}
where the factors $a_{1,2}$ and $b_{1,2}$ do not depend on $k_{\perp}^{2}$.
Ref.~\cite{egtfact} predicts that $b_{1}>b_{2}\geq1$. It is evident
that both Eq.~(\ref{r}) and Eq.~(\ref{rzot}) satisfy the restrictions
of Eq.~(\ref{rsmallk}), so we conclude that some more investigation
is required to constrain the form of the $R$ function.

\section{Discussion and outlook}

In this paper we have constructed the non-perturbative inputs for
the elastic gluon-hadron scattering amplitudes in the forward kinematic
region for both polarized and non-polarized hadrons. The optical theorem
allowed us to relate the invariant amplitudes to the gluon distributions
in the hadrons. The conventional approach is purely phenomenological
and constructs such inputs by matching with experimental data; in
contrast, we use the structure of the IR and UV divergences of the
factorization convolutions to determine the general requirements of
Eqs.~(\ref{mir},\ref{muv}). Imposing these conditions on the non-perturbative
inputs, we constructed the results for the Basic Factorization framework,
and then related them systematically to both the $K_{T}$- and Collinear
Factorization expressions. In the Basic Factorization framework, the
non-perturbative inputs consist of the invariant amplitudes $M_{U,S}$.
For simplicity we assumed $M_{U,S}$ had the same polarization structure
as for the perturbative case, but this is not obligatory. 

We then used the Resonance Model to suggest a form for the $Z$ factors
of Eq.~(\ref{musgen}), and assessed the criteria for which the resonance
factors were valid both within and outside of their resonance regions.
Starting from the Basic Factorization results, we could then extract
the corresponding results for non-perturbative inputs for the $K_{T}$-
and Collinear Factorizations. The inputs for $K_{T}$-Factorization
of Eqs.~(\ref{psi},\ref{tkt}) are of the resonance form, and we
can then use these to derive the expressions of Eq.~(\ref{phi})
for the input $\varphi$ in the Collinear Factorization framework.

\section{Acknowledgements}

We are grateful to A.~van Hameren and O.V.~Teryaev for useful discussions.
We acknowledge the hospitality of CERN, DESY, Fermilab, and UW-INT
where a portion of this work was performed. This work was partially
supported by the U.S. Department of Energy under grant DE-FG02-13ER41996,


\begin{thebibliography}{10}
\bibitem{szczur} A. Szczurek. Acta Phys.Polon.Supp. 8 (2015) 2, 483

\bibitem{dglap} G.~Altarelli and G.~Parisi, Nucl.~Phys.B126 (1977)
297; V.N.~Gribov and L.N.~Lipatov, Sov.~J.~Nucl.~Phys. 15 (1972)
438; L.N.Lipatov, Sov.~J.~Nucl.~Phys. 20 (1972) 95; Yu.L.~Dokshitzer,
Sov.~Phys.~JETP 46 (1977) 641.

\bibitem{egtg1sum} B.I.~Ermolaev, M.~Greco, S.I.~Troyan. Riv.Nuovo
Cim. 33 (2010) 57.

\bibitem{colfact} D.~Amati, R.~Petronzio, G.~Veneziano. Nucl.
Phys. B 140 (1978) 54; A.V.~Efremov, A.V.~Radyushkin. Teor.Mat.Fiz.
42 (1980) 147; Theor.Math.Phys.44 (1980)573; Teor.Mat.Fiz.44 (1980)17;
Phys.Lett.B63 (1976) 449; Lett.Nuovo Cim.19 (1977)83; S.~Libby, G.~Sterman.
Phys. Rev. D18 (1978) 3252. S.J.~Brodsky and G.P.~Lepage. Phys.
Lett. B 87 (1979) 359; Phys. Rev. D 22 (1980) 2157; J.C. Collins and
D.E. Soper. Nucl. Phys.B 193 (1981) 381; J.C. Collins and D.E. Soper.
Nucl. Phys.B 194 (1982) 445; J.C. Collins, D.E. Soper and G.~Sterman.
Nucl. Phys.B 250 (1985) 199. A.V.~Efremov and A.V.~Radyushkin. Report
JINR E2-80-521; Mod.Phys.Lett. A24 (2009) 2803.

\bibitem{bfkl} E.A. Kuraev, L.N. Lipatov and V.S. Fadin, Sov. Phys.
JETP 44, 443 (1976); E.A. Kuraev, L.N. Lipatov and V.S. Fadin, Sov.
Phys. JETP 45, 199 (1977); I.I. Balitsky and L.N. Lipatov, Sov. J.
Nucl. Phys. 28, 822 (1978).

\bibitem{ktfact} S.~Catani, M.~Ciafaloni, F.~Hautmann. Phys. Lett.
B 242 (1990) 97; Nucl.Phys.B366 (1991) 135.

\bibitem{hefact} J.C.~Collins, R.K.~Ellis. Nucl.Phys. B360 (1991)
3.

\bibitem{sud} V.V.~Sudakov. Sov. Phys. JETP 3(1956)65.

\bibitem{egtfact} B.I.~Ermolaev, M.~Greco, S.I.~Troyan. Eur.Phys.J.
C71 (2011) 1750; B.I.~Ermolaev, M.~Greco, S.I.~Troyan. Eur.Phys.J.
C72 (2012) 1953.

\bibitem{egtquark} B.I.~Ermolaev, M.~Greco, S.I.~Troyan. Eur.Phys.J.
C75 (2015)7, 306.

\bibitem{collinsrapid} J.C.~Collins. PoS LC2008 (2008) 028.

\bibitem{cheredrapid} I.O.~Cherednikov, N.G.~Stefanis. AIP Conf.Proc.
1105 (2009) 327; Int.J.Mod.Phys.Conf.Ser. 4 (2011) 135-145.

\bibitem{sterm} O.~Erdogan, G.~Sterman. Phys. Rev. D 91 (2015)
6, 065033.




\bibitem{fits} G.~Altarelli, R.D.~Ball, S.~Forte and G.~Ridolfi,
Nucl.~Phys.~B496 (1997) 337; Acta Phys. Polon. B29(1998)1145. E.~Leader,
A.V.~Sidorov and D.B.~Stamenov. Phys. Rev. D73 (2006) 034023; J.~Blumlein,
H.~Botcher. Nucl. Phys. B636 (2002) 225; M.~Hirai at al. Phys. Rev.
D69 (2004) 054021.

\bibitem{zotov} A.V.~Lipatov, G.I.~Lykasov, N.P,~Zotov. Phys.
Rev. D89. (2014) 014001.


\end{thebibliography}
\end{document}